\documentclass{aa}
\usepackage{graphicx}

\usepackage{times}

\begin{document}
\title{Polarimetric observations of OH masers in proto-planetary nebulae
}

\author{M. Szymczak  \inst{1} 
	\and E. G\'erard \inst{2} 
	}
\institute{
	  Toru\'n Centre for Astronomy, Nicolaus Copernicus 
          University, Gagarina 11, 87-100 Toru\'n, Poland 
\and      GEPI, UMR 8111, Observatoire de Paris, 5 place J. Janssen, 92195 Meudon Cedex, France
	  }

\offprints{M.Szymczak}

\date{Received 18 March 2004 / Accepted  28 April 2004 }

\abstract{The 1612 and 1667\,MHz OH maser lines have been measured in
all four Stokes parameters in 47 proto-planetary nebula (PPN)
candidates. Out of 42 objects detected, 40 and 34 are 1612 and
1667\,MHz emitters, respectively.  The spectral extent of the
1667\,MHz line overshoots that of the 1612\,MHz line in about 80\% of
the targets. 52\% and 26\% of the 1612 and 1667\,MHz sources,
respectively, show linear polarization in at least some features.
Circular polarization is more frequent, occurring in 78\% and 32\% of
sources of the respective OH lines. The percentage polarization is 
usually small ($<$15\%) reaching up to 50$-$80\% in a few sources. 
Features of linearly polarized emission are usually weak (0.5$-$4\,Jy) 
and narrow (0.3$-$0.5\,km\,s$^{-1}$). The strength of magnetic field inferred
from likely Zeeman pairs in two sources of a few mG is consistent with
values reported elsewhere for those classes of objects. An upper limit
of the electron density in the envelope of OH17.7$-$2.0 derived from
the difference in the position angle of polarization vectors for the
two OH lines is about 1\,cm$^{-3}$. Distinct profiles of polarization 
position angle at 1612 and 1667\,MHz are seen in about one third of 
the sources and strongly suggest that the envelopes are permeated with
structured magnetic fields. The geometry of the magnetic field is
implicated as an important cause of the depolarization found in
some PPN candidates. For the subset of targets which show axisymmetric
shells in the optical or radio images we found a dominance of magnetic
field components which are orthogonal to the long axis of the
nebulae. This finding supports the hypothesis that such bipolar
lobes are shaped by the magnetic field.
          }

\titlerunning{Polarimetric observations of OH masers in PPNe}
 
\authorrunning{M. Szymczak \& E. G\'erard}

\maketitle

\section{Introduction}
It has been recognized that at the end of their lives on the asymptotic 
giant branch (AGB), low- and intermediate-mass stars (1$-$8M$_{\sun}$) 
experience intense mass loss at rates of more than
$10^{-5}$M$_{\sun}$\,yr$^{-1}$ (Habing\,\cite{habing96}, for a review). 
This results in slowly expanding molecular circumstellar envelopes which 
can produce strong maser lines. During about 10$^3$\,yr after cessation of 
mass loss, stars evolve through a transitional proto-planetary nebula (PPN) 
phase into planetary nebulae (PNe). In the PPN phase an extensive dust-gas 
envelope can enshroud a cool ($\le 10^4$\,K) post-AGB stellar core while 
the photoionization of circumstellar matter is not yet initiated 
(Kwok\,\cite{kwok93}). It is thought that somewhere during the PPN phase, 
the envelope must depart from the high degree of spherical symmetry seen in 
most AGB stars to a variety of axisymmetric structures observed in most PNe
(Ueta et al.\,\cite{ueta00}). Possible causes of the axisymmetric morphology 
of such outflows are generally attributed to: 1) the interaction of a fast 
post-AGB stellar wind ($\ge 10^3$\,km\,s$^{-1}$) with a slow AGB wind with 
an equatorial density enhancement, 2) the effect of the gravitational field 
of a binary companion and 3) the dynamic effect of the progenitor magnetic 
field (Balick \& Frank\,\cite{balick02}, for a review).
 
The role of the magnetic field in the shaping of PPN has recently received 
a considerable attention. Chevalier \& Luo (\cite{chevalier94}) proposed 
the magnetized wind-blown bubble model in which initially weak toroidal
magnetic fields carried in a fast wind are strengthened and produce prolate 
and bipolar structures (e.g. Garcia-Segura et al.\,\cite{garcia99};
Matt et al.\,\cite{matt00}). Soker (\cite{soker02}) argued that the magnetic
field does not play a global role in the shaping of PPN but that the locally 
strong magnetic field may facilitate the maser amplification. 

The presence of magnetic fields around AGB and late type stars has been proved 
by maser polarization studies (e.g. Kemball \& Diamond\,\cite{kemball97}; 
Szymczak \& Cohen \cite{szymczak97}; Szymczak et al.\,\cite{szymczak98}, 
\cite{szymczak01}; Bains et al.\,\cite{bains03a}). OH maser emission appears to 
be the best tracer of magnetic fields in the outer circumstellar regions 
providing information on the direction, strength and orientation of the fields. 

OH masers have been found in several PPN candidates during the surveys of very 
cold IRAS sources (Likkel\,\cite{likkel89}; te Lintel Hekkert\,\cite{hekkert91};
Hu et al.\,\cite{hu94}; te Lintel Hekkert \& Chapman\,\cite{hekkert96}) and of 
radio continuum sources or young PNe (Zijlstra et al.\,\cite{zijlstra89}). 
Some PPN candidates with OH emission were identified on the basis of their 
irregular spectra with a velocity range over 50\,km\,s$^{-1}$ 
(Zijlstra et al.\,\cite{zijlstra01}). Polarization properties of OH masers 
associated with PPNe are poorly known. Zijlstra et al. (\cite{zijlstra89}) 
measured the circular polarization in 5 objects and a high percentage 
polarization ($>$80\%) was found in IRAS17150$-$3754 at 1612\,MHz. Another object 
with strongly circularly polarized emission in all OH lines was IRAS16342$-$3814 
(te Lintel Hekkert \& Chapman\,\cite{hekkert96}). Miranda et al. (\cite{miranda01}) 
reported that the PPN K3-35 shows strong circular polarization at 1667\,MHz. 
However, nothing is known about the linear polarization of OH masers in PPNe.

In this study we describe the results of a full polarization survey of OH masers 
in 47 PPN candidates. With the full polarization data we wish to better understand 
the properties of the circumstellar material of PPNe and to see if there is 
evidence that the magnetic field can facilitate the development of axisymmetric 
morphology. Our data should help to choose the best candidates for mapping 
the magnetic fields with the aperture synthesis technique. This paper provides 
the first evidence for the dominance of the magnetic field components perpendicular 
to the long axis of proto-planetary nebula.  


\section{Observations}
The 1612 and 1667\,MHz lines of OH were observed with the upgraded Nan\c{c}ay 
Radio Telescope (NRT) (van Driel et al.\,\cite{vandriel96}) in a series of observing 
runs between February 2002 and June 2003. At the OH frequencies, the half-power 
beam-width of the telescope is $3\farcm5$ in the E$-$W direction and $19\arcmin$ in 
the N$-$S direction. At $\delta = 0\degr$, the beam efficiency is 0.65 and the point 
source efficiency is 1.4\,K\,Jy$^{-1}$. The system temperature is about 35\,K. 

The new focal system of the NRT consists of a dual-reflector offset Gregorian 
configuration fed by a corrugated horn. The horn itself is followed by an ortho-mode 
transducer providing two linear orthogonal polarizations.  The feed box can be rotated, 
with a precision of 0\fdg5, at any angle between from $-90\degr$ to $+90\degr$ from 
a rest position of $0\degr$ (where the position angles of the electric vectors
are $45\degr$ and 135\degr). The maximum level of the cross-polarization lobe is 
$-$22\,dB. The linearly polarized signals are then up-converted, equalized in 
amplitude and phase and combined in a 7\,GHz hybrid to yield right hand (RHC) and 
left hand (LHC) circularly polarized signals. The isolation of opposite circular
polarizations is better than 20\,dB. The signals of the four RF channels (the two 
linear polarizations without hybridization and the two circular polarizations) are 
then down converted and the four IF channels transported to the laboratory. 
This system directly provides 3 of the 4 Stokes parameters namely $I$, $Q$ and $V$ 
while the fourth parameter $U$ is readily obtained by rotating the feed box by 45\degr
as is apparent from equations (2) and (3) below. The gain of each polarization channel 
was measured with a noise diode at the beginning of each 3\,min integration scan 
to $\sim$5\% in absolute value and to within $\sim$1\% in relative value.
 
A 8192 channel autocorrelator configured into eight banks of 1024 channels was used 
as spectrometer. The two OH transitions were simultaneously observed with a spectral 
resolution of 0.07 or 0.14\,km\,s$^{-1}$. For a few sources, with an OH maser spectral
extent greater than 60\,km\,s$^{-1}$, the data were taken with a resolution of 
0.28\,km\,s$^{-1}$. The radial velocities were measured with respect to the local 
standard of rest. The two orthogonal linear polarizations and two opposite circular 
polarizations of each OH line were observed at feed box positions of 0 and 45\degr.
 
The typical integration time for each feed box position was about 20\,min.  
The calibration of the instrumental polarization and of the polarization position 
angle was verified with regular observations of W3OH and OH17.7-2.0. The polarization 
characteristics of those secondary calibrators are known from interferometric observations
(Garcia-Barreto et al.\,\cite{garciabarreto88}; Bains et al.\,\cite{bains03a}). 
Moreover, the percentage of linear polarization and the position angle (PA, measured 
eastward from north) of the electric vector of the continuum source 3C286 were measured 
at 1400\,MHz and found to be 8.5$\pm$0.5\% and 33$\pm$0\fdg3, respectively (Colom, private
communication), in good agreement with published values (Bridle et al.\,\cite{bridle72}).

The four Stokes parameters were calculated for each velocity channel using the following 
equations:

\begin{equation}
I = S(0) + S(90) = S(RHC) + S(LHC)
\end{equation} 

\begin{equation}
Q = S(0) - S(90)
\end{equation}

\begin{equation}
U = S(45) - S(-45)
\end{equation}

\begin{equation}
V = S(RHC) - S(LHC), 
\end{equation}

\noindent
where $S($0$)$, $S($90$)$, $S($45$)$ and $S(-45)$ are the line flux density at PAs of 
0\degr, 90\degr, 45\degr \,and $-$45\degr, respectively. $S($RHC$)$ and $S($LHC$)$ 
are the line flux densities for right and left circular polarizations, respectively.  
The linearly polarized flux density $p=\sqrt{Q^2 + U^2}$, fractional linear 
polarization $m_{\rm L} = p/I$, fractional circular polarization $m_{\rm C} = V/I$ 
and polarization position angle $\chi = 0.5 * $tan$^{-1}(U/Q)$ were derived from 
the Stokes spectra. The uncertainties in the fractional linear polarization and 
the polarization position angle (McIntosh \& Predmore\,\cite{mcintosh93}) are

\begin{equation}
\sigma_{\rm m_L}= (\sqrt{p^2 \sigma_I^2/I^2 + \sigma_Q^2})/I
\end{equation}

\noindent
and

\begin{equation}
\sigma_{\chi}= \sigma_Q/\sqrt{2}p,
\end{equation}

\noindent
where $\sigma_I$, $\sigma_Q$ and $\sigma_p$ are the rms noises of the $I$, $Q$ and 
$p$ spectra, respectively. As discussed by Wardle and Kronberg (\cite{wardle74}), 
the non-Gaussian probability distribution of the polarized electric vector may lead 
to systematic errors when the signal-to-noise ratio (SNR) is small. We have neglected 
this effect since we only considered emission features with a SNR greater than 5.

A typical $\sigma_I$ was about 35\,mJy for 0.14\,km\,s$^{-1}$ spectral resolution.  
Baseline subtraction was done by frequency switching. This mode introduces an error 
of less than 0.6\% in the polarization parameters, as compared to the position switching
mode. No correction for ionospheric Faraday rotation was applied to the data. 
The absolute uncertainty of the OH polarization measurements is less than 7\%.

Each target was observed 3$-$5 times on an irregular basis to check the quality of 
polarization data rather than to study the variability which is beyond the scope of 
the paper. Consistent polarization parameter measurements were obtained for all objects 
at epochs spanning 3-15 months. Data for one epoch are presented and analyzed here
with the exception of section 5.3 where data from 2$-$5 observations are used to 
derive the polarization position angles. This procedure is of special value for sources 
with weak $p$ flux to improve the accuracy of polarization position angle estimates.


\begin{table*}
\caption {The sample of OH PPN candidates.}
\begin{tabular}{l l c c r r c r r}
\hline
\hline
IRAS & Other & Epoch & Spectral & \multicolumn{2}{c}{1612\,MHz} & & \multicolumn{2}{c}{1667\,MHz} \\
\cline{5-6} \cline{8-9}
name& name  & (JD$-$2450000) & resolution &  $\Delta V$ & $S_i$ &  & $\Delta V$ & $S_i$ \\
    &       &       &(km\,s$^{-1}$) &(km\,s$^{-1}$) & (Jy\,km\,s$^{-1}$) & & (km\,s$^{-1}$) & (Jy\,km\,s$^{-1}$) \\
\hline
06530$-$0213 &              & 2411 & 0.14  &         & $<$0.06  &&           & $<$0.05    \\ 
07331$+$0021 &              & 2461 & 0.07  & 10.2    & 11.6     && 4.5$^*$({\it 10.9})  & 0.07  \\ 
07399$-$1435 &OH231.8$+$4.2 & 2442 & 0.28  & 32.0    &  3.2     && 101.5     & 179.5      \\ 
08005$-$2356 &              & 2410 & 0.28  & 106.0   &  3.0     && 4.0$^*$({\it 111.1}) & 0.2  \\ 
16342$-$3814 &OH344.1$+$5.8 & 2758 & 0.14  & 132.0   & 33.6     && 117.0     & 26.3       \\ 
16559$-$2957 &              & 2432 & 0.14  &         & $<$0.14  &&           &$<$0.09     \\ 
17079$-$3844 &              & 2502 & 0.07  &         & $<$0.13  &&           &$<$0.11     \\ 
17103$-$3702 &NGC6302       & 2602 & 0.14  &21.0$^*$ &  5.6     &&10.0$^{*1}$&   0.5      \\ 
17150$-$3754 &OH349.4$-$0.2 & 2769 & 0.14  & 6.0$^*$ &  1.1     && 33.0$^*$  & 10.4       \\ 
17150$-$3224 &OH353.8$+$3.0 & 2718 & 0.14  & 2.8$^*$({\it 28.5}) &  4.2     && 31.5      & 53.9 \\ 
17233$-$2602 &OH0.1+5.1     & 2495 & 0.14  & 17.7    &  7.3     &&           &$<$0.11     \\ 
17253$-$2831 &              & 2370 & 0.14  & 31.8    & 31.6     &&           &$<$0.10     \\ 
17347$-$3139 &              & 2314 & 0.14  & 127.0   &  5.1     && 11.0$^2$  &0.3         \\ 
17371$-$2747 &              & 2502 & 0.14  & 18.2$^*$& 11.6     &&           &$<$0.10     \\ 
17375$-$2759 &              & 2818 & 0.14  & 15.6$^*$&  5.7     && 13.8$^*$  & 1.4        \\ 
17375$-$3000 &              & 2602 & 0.14  & 44.4    & 11.6     && 24.4      & 0.6        \\ 
17385$-$3332 &OH355.6$-$1.7 & 2530 & 0.28  & 25.7    & 18.5     && 22.4      & 0.4        \\ 
17393$-$2727 &OH0.9$+$1.3   & 2672 & 0.07  & 32.1    &199.2     && 8.2$^*$({\it 34.2})   & 1.3 \\ 
17393$-$3004 &              & 2821 & 0.07  & 62.5    & 30.7     && 38.0$^3$  & 2.9        \\ 
17404$-$2713 &OH1.2$+$1.3   & 2672 & 0.14  & 29.2    & 24.8     && 30.5      & 2.7        \\ 
17423$-$1755 &HEN3-1475     & 2327 & 0.28  & ({\it 24.3}) &  0.2  && 54.0    & 3.6        \\ 
17433$-$1750 &              & 2628 & 0.28  & 32.0    &  3.4     && 31.8      & 0.3        \\ 
17436$+$5003 &HD161796      & 2383 & 0.07  &         &$<$0.12   &&           &$<$0.10     \\ 
17443$-$2949 &              & 2820 & 0.14  & 24.8$^4$&  3.5     && 31.5$^5$  & 3.9        \\ 
17516$-$2525 &              & 2432 & 0.14  & 37.2    &  9.4     && 37.2      & 0.8        \\ 
17579$-$3121 &              & 2683 & 0.14  & 24.1    & 47.6     && 49.5      & 0.9        \\ 
18016$-$2743 &              & 2419 & 0.14  & 24.8    &  3.5     &&           &$<$0.12     \\ 
18025$-$3906 &              & 2558 & 0.14  & 51.6    & 41.0     &&           &$<$0.23     \\ 
18052$-$2016 &OH10.1$-$0.1  & 2434 & 0.14  & 63.0    & 17.0     &&           &$<$0.14     \\ 
18071$-$1727 &OH12.8$+$0.9  & 2421 & 0.07  & 22.3    &  8.1     && 23.7      & 1.2        \\ 
18091$-$1815 &              & 2436 & 0.28  & 32.9    & 16.8     && 34.0$^6$  & 1.1        \\ 
18095$+$2704 &OH53.8$+$20.2 & 2440 & 0.14  & 20.8    &  0.4     &&           & $<$0.11    \\ 
18105$-$1935 &              & 2442 & 0.14  & 21.8    &  8.5     && 25.5      & 0.6        \\ 
18135$-$1456 &OH15.7$+$0.8  & 2694 & 0.07  & 31.3    &156.6     && 32.2      & 13.5       \\ 
18266$-$1239 &OH19.2$-$1.0  & 2499 & 0.14  & 37.7    & 76.2     && 38.0      &  0.6       \\ 
18276$-$1431 &OH17.7$-$2.0  & 2326 & 0.14  & 28.0    &299.2     && 28.4      & 13.2       \\ 
18450$-$0148 &W43A          & 2777 & 0.14  & 16.2    & 37.9     && 19.2      &  8.6       \\ 
18491$-$0207 &              & 2421 & 0.28  &         &$<$0.20   &&132.0      & 37.4       \\ 
18596$+$0315 &OH37.1$-$0.8  & 2430 & 0.14  & 30.1    & 39.0     && 31.5      &  1.2       \\ 
19067$+$0811 &OH42.3$-$0.1  & 2694 & 0.14  & 35.3    & 74.3     && 3$^*$({\it 35.5})  &  0.2   \\ 
19114$+$0002 &HD179821      & 2809 & 0.28  & 55.3    &297.5     && 58.8      &184.9       \\ 
19127$+$1717 &              & 2438 & 0.14  &         &$<$0.17   &&           &$<$0.12     \\ 
19219$+$0947 & VY2-2        & 2775 & 0.07  & 8.5$^*$ & 19.5     &&           &$<$0.10     \\ 
19255$+$2123 & K3-35        & 2420 & 0.07  & 25.4    &  3.4     && 5.6$^*$   & 0.2        \\ 
19343$+$2926 & M1-92        & 2421 & 0.14  & 4.0$^*$({\it 45.8}) &  0.4     && 46.0   & 7.9   \\ 
22036$+$5306 &              & 2335 & 0.28  & 60.7    & 14.7     && 66.5      & 9.0        \\ 
23321$+$6545 &              & 2447 & 0.14  &         &$<$0.14   && 20.7      & 0.3        \\ 
\hline
\end{tabular}

$^*$ width of single feature or complex;
Absorption features: $^1$$-$0.1\,Jy at $-$37.8\,km\,s$^{-1}$, 
$^2$$-$0.2\,Jy at $-$7.2\,km\,s$^{-1}$, 
$^3$$-$0.4\,Jy at 1.2\,km\,s$^{-1}$, 
$^4$$-$0.1\,Jy at 4.8\,km\,s$^{-1}$, 
$^5$$-$0.1\,Jy at 4.8\,km\,s$^{-1}$,
$^6$$-$0.1\,Jy at 30.7\,km\,s$^{-1}$ \\
\end{table*}


\section{The sample}
We selected 47 target sources from several surveys of OH masers of IRAS sources 
with dust temperatures between $\sim$100\,K and $\sim$200\,K.  This range is 
typical for late AGB stars and early post-AGB heavily obscured by thick 
circumstellar envelopes only detectable in the infrared, as well as for post-AGB 
and young PN still associated with the infrared emission but with optically visible
central stars of spectral types from M to B (Likkel\,\cite{likkel89};
Zijlstra et al.\,\cite{zijlstra89}; te Lintel Hekkert\,\cite{hekkert91}; 
Hu et al.\,\cite{hu94}; te Lintel Hekkert \& Chapman\,\cite{hekkert96}; 
Zijlstra et al.\,\cite{zijlstra01}). OH maser emission from those objects at one 
or more transitions is usually over 50\,km\,s$^{-1}$ wide and significantly 
different from the conventional standard twin-peaked AGB star OH 1612-MHz maser
profile. Sometimes the red-shifted peaks are weak or absent. Some objects with 
a moderate OH velocity range of 20-30\,km\,s$^{-1}$ are included because they 
have been revealed as bipolar nebulae by radio observations (e.g. W43A, K3-35). 
Our sample also contains some objects detected as OH masers during the surveys 
of known young PN or radio continuum sources (Zijlstra et al.\,\cite{zijlstra89}).
We did not include PPN candidates lying in the directions of strong background OH
confusion where the genuine OH emission of the source cannot be unambiguously 
resolved with the telescope beam. The well known OH supergiant stars with high 
outflow velocities are excluded from our list. The sample is confined to the objects 
with $\delta > -$39\degr.


\section{Results}
Table 1 lists the objects observed and contains velocity ranges ($\Delta V$) 
and integrated flux densities ($S_i$) at 1612 and 1667\,MHz. 
For sources with only the blue- or red-shifted emission detected the measured
values of $\Delta V$ are denoted by asterisks while the presumed values of 
$\Delta V$, discussed below in section 4.6, are given in italics. 
For the non-detections we list the 3$\sigma$ noise levels.

Figure 1 shows the 1612\,MHz polarization spectra of OH17.7$-$2.0. 
Circularly and linearly polarized emission was detected in the blue- and 
red-shifted parts of the spectrum. The percentage of linear and circular
polarization was generally lower than 8\% and $\pm$15\%, respectively. 
We note that changes in the PA of linear polarization across the spectrum 
closely follow the pattern observed with MERLIN (Bains et al.\,\cite{bains03a}). 
This result demonstrates that the NRT spectrum provides a reliable measure of 
the magnetic field orientation in this circumstellar envelope the structure of 
which was resolved with a 0\farcs2 beam. Furthermore, it suggests that the 
field orientation is stable over a period of about 2.5 years. Because our data 
were taken with a spectral resolution of a factor of 2.6 higher than that of 
MERLIN but with a sensitivity about a factor of 4.5 lower, the PA profile is
noisier than that deduced from Bains et al.\,(\cite{bains03a}).

The polarization spectra for the remaining sources for which the $p$
spectra contain at least one feature with SNR$>$5 are shown in Fig. A1.

\begin{figure}
   \resizebox{\hsize}{!}{\includegraphics{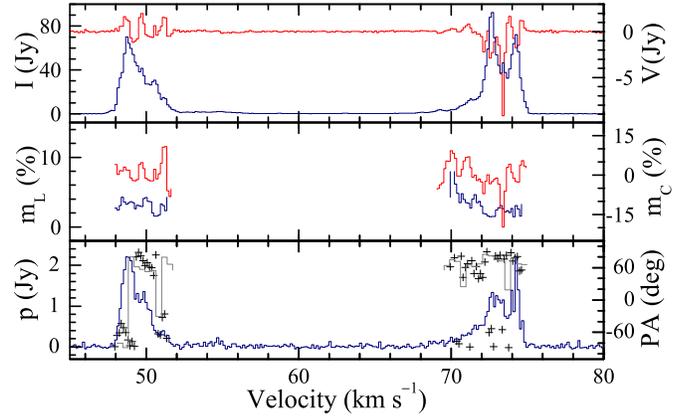}}
   \caption{Polarization spectra for the 1612\,MHz transition in 
   IRAS18276$-$1431 (OH17.7$-$2.0) taken on 2002 February 20.    
   Upper panel: the total flux ($I$ Stokes) (blue line) and 
   the circularly polarized flux ($V$ Stokes) (red line);
   middle panel: the percentage linear polarization ($m_{\rm L}$) (blue
   line) and the percentage circular polarization 
   ($m_{\rm C}$) (red line); lower panel: the linearly polarized flux 
   ($p =\sqrt{Q^2+U^2}$ Stokes) (blue line) and the position angle 
   (PA) of the polarization vector (crosses). The grey line in 
   the lower panel shows the flux-weighted PA of polarization vector 
   observed with MERLIN on 1999 May 19 (Bains et al.\,\cite{bains03a}).}
   \label{fig1}
\end{figure}

\subsection{Polarization statistics}
Among the 42 objects detected, there are 32 objects with polarized 
(circular and/or linear) emission in at least one OH transition (Table 2). 
23 sources show linearly polarized emission in either or both of the OH 
lines. Linearly polarized emission is always associated with circularly 
polarized emission. 10 out of 31 polarized objects at 1612\,MHz show only 
circular polarization (Table 2). There is a higher probability of detecting 
linear polarization in the 1612\,MHz line (21/40) than in the 1667\,MHz 
line (9/34) (Tables 1 and 2).

The distribution of source counts versus integrated flux density (Fig. 2) 
demonstrates that polarized emission preferentially occurs in objects with 
strong emission; at 1612\,MHz the average values of $S_i$ in polarized and 
non-polarized sources are 19.4 and 2.3\,Jy\,km\,s$^{-1}$, respectively, 
whereas at 1667\,MHz the corresponding values are 16.7 and 1.1\,Jy\,km\,s$^{-1}$. 
Plot of $|V|$ and $p$ fluxes versus the total flux $I$ of the strongest peak 
for the whole sample (Fig.3) clearly shows that only one source with flux 
less than 0.8\,Jy had detectable polarization. Below a flux density of about
2\,Jy the polarized source counts are likely to be incomplete.

\begin{figure}
   \resizebox{\hsize}{!}{\includegraphics{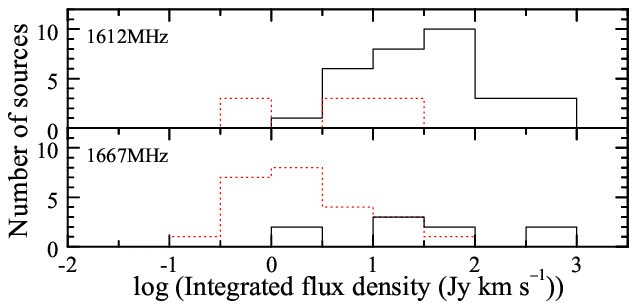}}
   \caption{Histograms of the integrated flux density for sources with polarized 
            (solid line) and non-polarized (dashed line) OH emission.}
   \label{fig2}
\end{figure}

\begin{figure}
   \resizebox{\hsize}{!}{\includegraphics{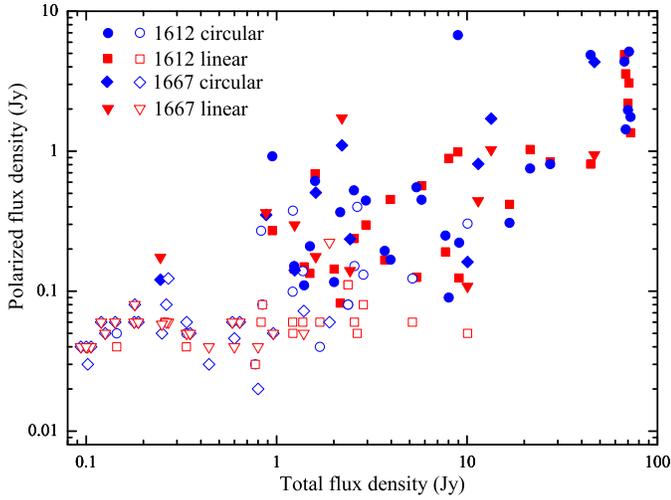}}
   \caption{Polarized flux densities ($|V|$ and $p$) plotted against
            the total $I$ flux densities. The filled symbols show
            sources with linearly (red) and circularly (blue)
            polarized features.  The open symbols show sources with
            only an upper limit for polarized emission.}
   \label{fig3}
\end{figure}

\subsection{Strongly polarized sources}
There are 7 OH PPN candidates in the sample with strongly polarized features 
(Table 2 and Fig.A1). The most prominent is IRAS16342$-$3814 containing, in 
the blue-shifted part of the 1612\,MHz spectrum, several polarized features 
with $|m_{\rm C}|$$>$70\% and $m_{\rm L}$=14\%. The 1612\,MHz features of 
IRAS19255$+$2123 have even higher polarization percentages 
($|m_{\rm C}|$$>$80\%, $m_{\rm L}$=40\%). IRAS08005$-$2356 shows considerable 
polarization at 1612\,MHz ($|m_{\rm C}|$=37\%, $m_{\rm L}$=51\%) near 
$-$0.3\,km\,s$^{-1}$. In the 1612\,MHz line strong circular polarization 
($|m_{\rm C}|$=84\%) without detectable linear polarization is seen in 
IRAS17150$-$3754, while in IRAS19067$+$0811 strong linear polarization of 
$m_{\rm L}$=70\% is associated with only weak circular polarization 
($m_{\rm C}$=2.1\%). The 1667\,MHz emission of IRAS18091$-$1815 and
IRAS18276$-$1431 is characterized by considerable linear polarization
($m_{\rm L}$=54\%). We note that these features with a high degree of
polarization are usually weak in total intensity ($<$ 2\,Jy).

\subsection{Evidence for complex magnetic fields}
The data for OH17.7$-$2.0 (Fig.1) indicate that the measurements of
the position angles of polarization vectors are fully consistent with
those obtained using an interferometric array. In this source, a
large-scale regular magnetic field structure was revealed by Bains et
al.\,(\cite{bains03a}). It is possible that sources with considerable 
changes in $\chi$ across the $p$ profile have structured magnetic fields.  
In our sample the source IRAS17393$-$2727 appears as the best candidate 
to map a complex magnetic field; a $p$ flux density higher than 1\,Jy is 
seen over a velocity extent of greater than 6\,km\,s$^{-1}$, $\chi$ varies 
in a characteristic way from $-90\degr$ to $90\degr$ (Fig. A1). Two further 
objects, IRAS07399$-$1435 at 1667\,MHz and IRAS19067$+$0811 at 1612\,MHz, 
with linearly polarized flux densities of about 1\,Jy, could also have 
structured fields. Evidence for large variations in the PA of polarization 
vectors across the $p$ profile is seen for IRAS16342$-$3814, IRAS17150$-$3224 
and IRAS1404$-$2713 at 1612\,MHz. In these targets, however, the linearly 
polarized flux is quite weak (usually less than 0.5\,Jy) and the width of 
$p$ features is as narrow as 0.3\,km\,s$^{-1}$.

\begin{table*}
\caption {The strongest polarized OH features in PPN candidates.}
\begin{tabular}{l r r r r r r }
\hline
\hline
IRAS name & Line   & $V_c$  & $m_{\rm C}$($\sigma$) & $V_l$ & $m_{\rm L}$($\sigma$) & $\chi$($\sigma$) \\
          & (MHz)  &(km\,s$^{-1}$) &    (\%)        & (km\,s$^{-1}$) & (\%) & (\degr) \\
\hline
07331$+$0021   & 1612 &  32.3     & 5.0(1.1) & 32.5   & 12.3(0.4) & $-$31.6(1.6)  \\ 
07399$-$1435   & 1667 &  19.0  &$-$2.7(0.2) & 20.4   & 10.7(0.4) & 54.2(2.4)\\
08005$-$2356   & 1612 & $-$0.3   &$-$37.0(1.5) &$-$0.4  & 51.3(5.1) & 30.5(5.7) \\ 
16342$-$3814   & 1612 & $-$7.3   &$-$70.8(2.1) &$-$7.5  & 14.2(1.9) &$-$43.8(10.7) \\
               & 1667 &  9.8 & 13.8(1.6)&$-$4.2 & 26.5(4.0) &$-$61.8(13.0)\\
17150$-$3754   & 1612 & $-$125.3 &$-$84.2(6.3) &        &       &        \\ 
17150$-$3224   & 1612 & 25.0     &$-$5.2(0.8)  & 24.9   & 9.0(1.0)  & 79.8(2.4)\\
               & 1667 & 26.0  & 8.6(1.4)& 25.9   & 4.2(0.4)  &$-$68.2(5.6)\\ 
17253$-$2831   & 1612 & $-$71.6  &$-$2.9(0.3)  &       &          &    \\ 
17347$-$3139   & 1612 & $-$23.1  & 11.9(2.4)   &        &       &       \\ 
17375$-$2759   & 1612 & 22.2     &$-$7.1(2.6) & 23.8   & 16.1(2.0)  &$-$17.5(8.3)  \\ 
17375$-$3000   & 1612 & $-$35.6  &$-$9.5(1.2) &        &       &       \\ 
17385$-$3332   & 1612 & $-$245.6 &$-$7.0(2.3) &$-$224.3& 4.1(1.1)   & 19.6(12.1)    \\ 
17393$-$2727   & 1612 & $-$120.5 & 12.9(0.8)  &$-$123.5& 4.2(0.2)  &$-$60.8(3.3) \\
               & 1667 & $-$124.5&$-$13.5(3.2)&$-$124.5& 24.3(4.3)&6.6(13.8)\\
17393$-$3004   & 1612 & 16.1     & $-$29.6(2.6)&$-$27.4 & 11.4(0.5) &$-$23.3(5.5) \\
               & 1667 &$-$22.9&$-$38.7(20.5)&   &          &          \\
17404$-$2713   & 1612 & 30.7     & $-$6.6(0.4) & 31.7   & 5.9(0.9) & 57.0(4.2)  \\
               & 1667 & 54.8  &$-$24.9(5.5)& 55.0 & 19.6(6.2) & $-$1.3(4.8)\\
17443$-$2949   & 1612 & $-$15.4  & $-$12.8(1.0)    &        &       &       \\ 
17579$-$3121   & 1612 & $-$0.2   &$-$9.3(1.1) & 22.3   & 3.0(0.1) &$-$34.5(5.5)  \\ 
18016$-$2743   & 1612 & 61.4     & 12.0(2.8)  & 61.4   & 15.9(2.7) & 0.2(10.6)   \\ 
18025$-$3906   & 1612 & $-$129.1 & 23.1(3.0)  &        &           &             \\ 
18052$-$2016   & 1612 & 79.5     & 5.7(0.4)    &        &       &       \\ 
18071$-$1727   & 1612 & 17.6     & 17.0(0.7)   &        &       &       \\
18091$-$1815   & 1612 & 10.6     & $-$5.4(7.0) &        &       &   \\
               & 1667 &  8.3  & 32.4(4.7) & 8.3    & 53.8(3.4)  &$-$49.9(16.0)\\ 
18105$-$1935   & 1612 & 8.5      & $-$7.0(0.4) &        &       &        \\ 
18135$-$1456   & 1612 & 13.5     & 3.8(0.2) & 13.1 & 5.1(0.2) & $-$21.2(3.8)\\
               & 1667 &  13.9 & $-$7.6(0.6)&  &  &  \\
18266$-$1239   & 1612 & 57.5     &$-$2.6(0.5) & 57.8   & 5.6(0.5)& 71.9(4.3)  \\ 
18276$-$1431   & 1612 & 73.3     &$-$16.1(3.6)& 48.8 & 3.5(0.1) &$-$69.8(6.0)\\
               & 1667 & 74.3 &$-$15.4(1.3)& 72.6&72.0(9.0)& 80.4(7.0)\\ 
18450$-$0148   & 1612 & 40.5     & 11.3(1.8) & 40.3 &  2.4(0.2) & 63.3(5.2) \\
               & 1667 & 40.6 & 21.1(1.2) & 26.4 & 19.0(2.4) & 77.4(6.2) \\ 
18596$+$0315   & 1612 & 99.4     &$-$11.0(2.4) & 76.9   & 4.7(0.5) &$-$3.1(4.4)   \\ 
19067$+$0811   & 1612 & 75.6     & 2.1(0.6)  & 76.5  & 69.9(18.9) & 87.9(4.3) \\
19114$+$0002   & 1612 & 123.8    &$-$6.5(0.2)&123.7 & 7.5(0.1)  &$-$88.4(0.5)\\
               & 1612 & 124.9 &$-$8.6(0.1) &124.9 & 1.8(0.1) &$-$86.9(2.6)\\
19219$+$0947   & 1612 & $-$61.9  & 6.6(0.8) &$-$62.1 & 4.0(0.4) & 3.9(10.0)   \\ 
19255$+$2123   & 1612 & 21.2     &$-$83.4(12.3)  & 21.3 & 39.9(10.1)& $-$57.2(22.7) \\ 
22036$+$5306   & 1612 & $-$72.5  & 6.1(16.3) &$-$63.5 & 23.2(1.1) &$-$74.1(1.1)  \\
\hline
\end{tabular}
\end{table*}

\subsection{Zeeman splitting}
At 1612\,MHz we found S-shaped features in the $V$ spectra of IRAS07331$+$0021 
and IRAS18266$-$1239 (Fig. 4). It is possible that they arise due to the Zeeman 
effect, which could be confirmed by high angular resolution observations.  
Assuming that for the $\sigma$ components an average velocity splitting of 
0.236\,km\,s$^{-1}$ is produced by a magnetic field of strength of 1\,mG, the 
corresponding strength of magnetic field along the line of sight is 
$-$1.07$\pm$0.24\,mG for IRAS07331$+$0021 and $-$2.57$\pm$0.14\,mG for
IRAS18266$-$1239. A negative sign indicates a field pointing towards the observer.
These estimates are consistent with the field strengths measured in AGB stars 
and other PPN candidates (Szymczak et al.\,\cite{szymczak98}; 
Bains et al.\,\cite{bains03a}).

\begin{figure}
   \resizebox{\hsize}{!}{\includegraphics{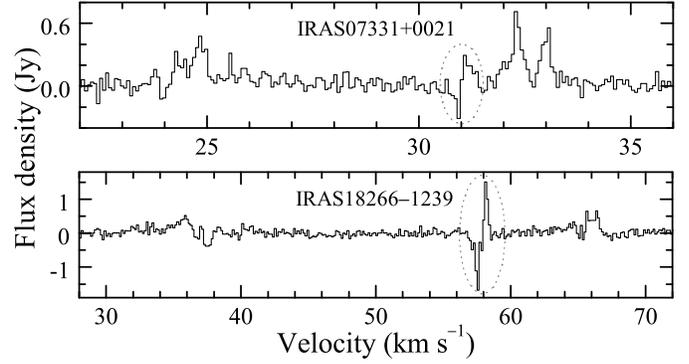}}
   \caption{$V$ Stokes 1612\,MHz OH spectra of PPN candidates with possible 
            Zeeman splitting marked by dashed ovals.}
   \label{fig4}
\end{figure}

\begin{figure}
   \resizebox{\hsize}{!}{\includegraphics{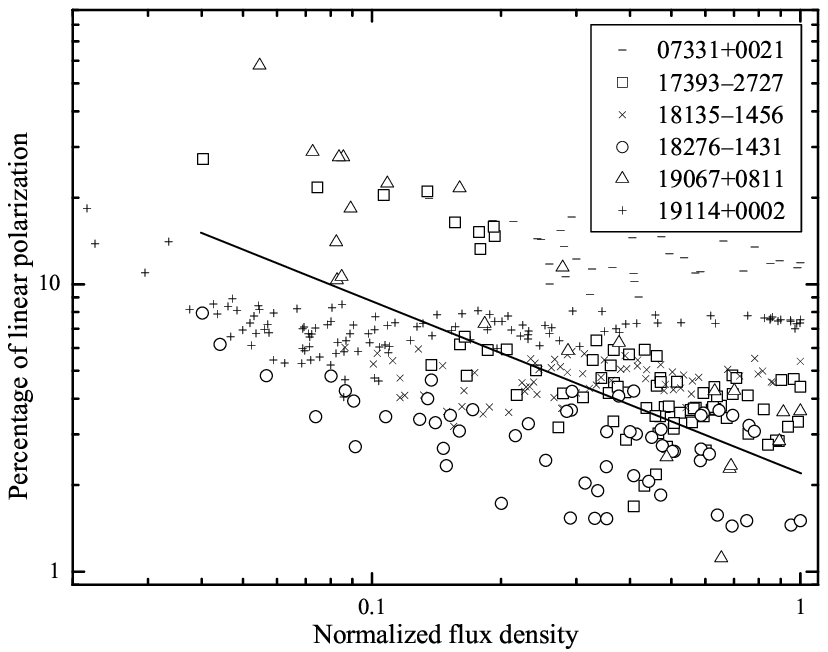}}
   \caption{Percentage of linear polarization versus the normalized total flux at 1612\,MHz
            for sources with $p$ flux higher than 5$\sigma$ seen in more than 35 spectral channels. 
            Each point represents data for channel of width 0.07 or 0.14\,km\,s$^{-1}$, 
            and 0.28\,km\,s$^{-1}$ for IRAS19114+0002 (see Table 1). The line of slope of
            $-$0.57 is the best fit to the data marked by open symbols. Note the absence of
            decreasing trends for targets marked by small symbols.} 
\label{fig5}
\end{figure}

\subsection{Depolarization}
A reduced degree of polarization for bright $I$ features is seen in some objects 
in the sample. Figure 5 shows the percentage linear polarization as a function 
of the normalized total flux density for 6 objects with $p$ flux seen in at least 
35 spectral channels at 1612\,MHz. Decreasing trends are clearly visible for 
IRAS17393$-$2727, IRAS18276$-$1431 and IRAS19067$+$0811. The depolarization effect 
can be described by a power law of the form $m_{\rm L}= A I^{\alpha}$ where 
$\alpha=-0.57\pm0.05$ effectively characterizes all the data for these three sources. 
This trend cannot be accounted for by the uncertainties in the fractional 
polarization. Observations of these sources with different spectral resolutions 
(usually 0.07 or 0.14\,km\,s$^{-1}$) have no significant influence on the decreasing
trend. Sources IRAS07331+0021, IRAS18135$-$1456 and IRAS19114+0002 do not show 
diminished polarization in spectral channels with high total intensities (Fig.5).  
For the remaining sources, the linearly polarized emission was detected in too few 
(3$-$25) channels to make a meaningful analysis of depolarization.

In looking for possible causes of depolarization it is worth noting that the extremely 
low dispersion ($1\fdg1 - 2\fdg3$) in the position angle of polarization vectors among 
the sources which do not show the depolarization effect, compared with $4\fdg3 - 12\fdg5$ 
dispersion among the objects with depolarization (Fig. A1). This suggests that the
depolarization effect depends on the global magnetic field structure in the envelope; 
the directions of the projected field in the sources without depolarization are well 
aligned along specific PAs, whereas in the sources with depolarization the field geometry 
appears to be much more complex.

In the case of circularly polarized emission we note strong depolarization in 
IRAS17393$-$2727 and IRAS19067$-$0811. No depolarization is seen in IRAS18276$-$1431 
and in the other three sources not showing linear depolarization. This suggests that
mechanisms for depolarization may be common to linear and circular emission.

\subsection{Overshoot of 1667\,MHz spectral extent}
There are 20 objects with well detected blue- and red-shifted parts of spectra 
at 1612 and 1667\,MHz (Table 1) and in 15 of them the spectral extent of the 
1667\,MHz line is by at least 0.2\,km\,s$^{-1}$ larger than that of the 1612\,MHz line.  
All 7 objects with only the blue-shifted or red-shifted emission at one maser line 
(estimates of the presumed spectral extent are given in Table 1 as italic) but with
the complete profile at other maser line also exhibit the overshoot of the 1667\,MHz 
line relative to the 1612\,MHz maser extent. The absolute value of overshoot usually 
ranges from 0.2 to 6\,km\,s$^{-1}$. The distribution of the ratio of the velocity extent
at 1612 and 1667\,MHz is shown in Figure 6. The overshoot effect is seen in 81\% (22/27) 
of the objects in the sample.

The overshoot phenomenon is less prominent and less frequent (27$-$29\%) in OH/IR objects 
(Dickinson \& Turner\,\cite{dickinson91};Sivagnanam et al.\,\cite{sivagnanam89}). 
Although there is no unique mechanism for overshoot (Sivagnanam \& David\,\cite{sivagnanam99}) 
a departure from spherical symmetry in the envelope is a plausible cause; most PPN candidates 
observed at high angular resolution show non-spherical geometry (Zijlstra et al.\,\cite{zijlstra01}).

\begin{figure}
   \resizebox{\hsize}{!}{\includegraphics{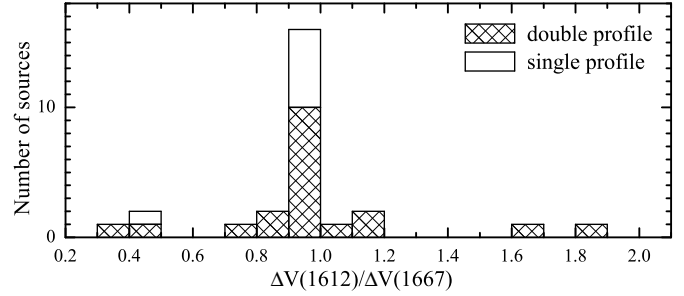}}
   \caption{Histogram of the ratio of the spectral maser extents at 1612 and 1667\,MHz.} 
\label{fig6}
\end{figure}

\subsection{Absorption features}
We found 5 objects with absorption features, usually at 1667\,MHz (Table 1).  
All these are inner Galaxy low latitude sources (349\degr$<l<$ 13\degr, $|b|<$1.1\degr). 
In IRAS17347$-$3139, IRAS17393$-$3004 and IRAS17443$-$2949 the velocities close to zero 
of the absorption features strongly suggest that they are interstellar in nature. 
OH maps by Boyce \& Cohen (\cite{boyce94}) confirm this unambiguously for the latter 
two objects. The absorption feature of IRAS18091$-$1815 near 30.7\,km\,s$^{-1}$ appears 
in the red-shifted part of the OH spectrum and is possibly not related to the source.  
In IRAS17103$-$3702 the 1667\,MHz absorption near $-$37.8\,km\,s$^{-1}$ 
(see also Payne et al.\,\cite{payne88} and Zijlstra et al.\,\cite{zijlstra89}) 
coincides in velocity with the edge of the 1612\,MHz maser emission. As only the 
blue-shifted emission is observed it is likely that the absorption arises in front of 
the ionized region of the nebula. This OH PPN object possibly represents an older phase 
of evolution. We conclude that in the studied sources, the OH absorption features are 
generally interstellar in origin.

\section{Discussion}
The present systematic data reveal an OH polarization pattern in PPN candidates which 
is quite different from that described in the literature for late-type stars.  
Out of 40 targets detected at 1612\,MHz, 21 and 31 were found to show linearly and 
circularly polarized features, respectively. Early observations did not provide evidence 
for polarization in many PPNe, with the exception of a few stars during flare activities, 
(Cohen\,\cite{cohen89}, for a review). Considerable circular polarization was found in 
supergiants (Cohen et al.\,\cite{cohen87}) and OH/IR stars (Zell \& Fix\,\cite{zell91}) 
when observed with high sensitivity and spectral resolution. 
Zijlstra et al.\,(\cite{zijlstra89}) measured circular polarization in 5 PPN candidates 
and polarized emission was detected in IRAS17150$-$3754 and IRAS17393$-$2727.

Out of 34 1667\,MHz objects, 9 exhibited linearly polarized emission. This detection 
ratio is about 2.5 times higher than that reported for Mira-type stars 
(Olnon et al.\,\cite{olnon80}; Claussen \& Fix\,\cite{claussen82}). Because linearly 
polarized features are generally weak (0.5$-$4\,Jy) and narrow (0.3$-$0.5\,km\,s$^{-1}$)
their only occasional detection in the past seems to be partly an effect of insufficient 
sensitivity and spectral resolution.

We note that at the two OH frequencies studied, linearly polarized features are always 
associated with circularly polarized features, but circular polarization can be present 
without associated linear polarization. A similar correlation has been reported for AGB 
stars (Claussen \& Fix\,\cite{claussen82}). Appearance of circular polarization alone
suggests that the effect of magnetic beaming (Gray \& Field\,\cite{gray95}) can work
in the PPN objects.

\subsection{Upper limit for electron density} 
In IRAS18276$-$1431 and IRAS19114$+$0002 linearly polarized features in the red-shifted 
parts of spectra at the two transitions were observed at the same velocities (Fig. 1 and 
Fig. A1). Strong enough features ($>$0.5\,Jy) at 1612 and 1667\,MHz were identified in
IRAS18276$-$1431 near 72.8 and 74.4\,km\,s$^{-1}$ and in IRAS19114$+$0002 near 
124.9\,km\,s$^{-1}$. Assuming that both lines emerge in the same volume of gas, any 
difference in $\chi$ between the two lines would probably be due to Faraday rotation 
over the envelope path length. No significant differences in PA exist within errors of 
9\fdg9 and 12\fdg4 for IRAS18276$-$1431 and IRAS19114$+$0002, respectively. As the radii 
of OH envelopes at the transitions of interest were determined in both targets 
(Bains et al.\,\cite{bains03a}; Gledhill et al.\,\cite{gledhill01}) and the strength of 
the magnetic field along the line of sight was measured in the first object (4.6\,mG) 
we can estimate an upper limit of the mean electron density in the envelope from 
the expression for Faraday rotation (Garcia-Barreto et al.\,\cite{garciabarreto88}):

\begin{equation}
\phi =0\fdg5 n_e B_{||} L \lambda^2,
\end{equation}

\noindent
where $n_e$ is the electron density in cm$^{-3}$, $B_{||}$ is the strength of magnetic 
field along the line of sight in mG, $L$ is the depth of the maser region in units 
of 10$^{15}$cm and $\lambda$ is the wavelength of the transition in units 18\,cm. 
For an adopted distance of 2\,kpc to OH17.7$-$2.0 (IRAS18276$-$1431) the depth of 
the maser region is comparable with the diameter of the envelope of 5.1$\times10^{16}$cm 
(Bains et al.\,\cite{bains03a}). The radiation comes from the far side of the envelope. 
The above observational parameters of OH17.7$-$2.0 imply an upper limit for
$n_e$ of 1.2\,cm$^{-3}$. Such a low value suggests largely neutral matter in the envelope.

\subsection{Magnetic fields in PPNe}
Our polarimetric observations revealed a great variety of polarization properties of 
OH maser lines in PPNe. IRAS19255$+$2123 is a good example to illustrate the diversity of 
polarization properties. The 1612\,MHz emission is largely unpolarized with the exception 
of the feature near 21.2\,km\,s$^{-1}$ which shows elliptical polarization
($m_{\rm C} =-$83\%, $m_{\rm L}$=40\%). We failed to detect any polarized emission at 
1667\,MHz. Considerable circular polarization at 1665\,MHz was reported by Miranda et al. 
(\cite{miranda01}). As the emission from both main lines is located in different parts of 
the envelope from the 1612-MHz emission it is likely that their polarization parameters 
are due to the local magnetic field strength and orientation. The 1612\,MHz data imply that 
the PA of the projected magnetic field differs by 53\degr from the PA of the bipolar lobes
observed in radio continuum (Miranda et al.\,\cite{miranda01}). This result partly supports 
the theoretical models of outflows in PNe collimated by toroidal magnetic fields 
(Garcia-Segura et al.\,\cite{garcia99}; Matt et al.\,\cite{matt00}). However, the small
number of polarized OH features in this source prevents the study of the magnetic field 
orientation with high angular resolution in order to verify the existence of a magnetized
torus postulated by Miranda et al. (\cite{miranda01}).

Although the magnetic field of 1-2\,mG tentatively found in two PPN objects needs further 
confirmation by high angular resolution studies, its strength is consistent with that 
measured with MERLIN in IRAS18276$-$1431 (Bains et al.\,\cite{bains03a}) as well as in AGB
stars (Szymczak et al.\,\cite{szymczak98}). A simple consideration suggests that the magnetic 
field of the order of a few mG can shape the outflow in this object 
(Bains et al.\,\cite{bains03a}). They document how IRAS18276$-$1431 has a magnetic field 
with large-scale well ordered structure, the geometry of which is consistent with 
a stretched dipole. The signature of a structured field is seen in the polarization vector 
PA profile (Fig.1). Several sources in our sample, most prominently IRAS17393$-$2727, show 
clear evidence for a more complex magnetic field.  Among the promising candidates is
IRAS19114$+$0002 which shows broad linearly polarized features at red-shifted velocities 
at both lines and much weaker blue-shifted features at 1612\,MHz.  The strongest linearly 
polarized features, near 124\,km\,s$^{-1}$, are 2 and 7\% of the total peak flux densities
at 1667 and 1612\,MHz, respectively. MERLIN observations of both lines failed to detect any 
polarized emission above the noise level (Gledhill et al.\,\cite{gledhill01}) possibly due 
to insufficient spectral resolution. Our data show very small and smooth changes of 
the polarization position angles for the two lines, suggesting a magnetic field aligned 
along PA$\approx$5\degr. The distribution of the 1667\,MHz emission is elongated along PA=15\degr
(Gledhill et al.\,\cite{gledhill01}); suggesting that the magnetic field is well aligned 
approximately parallel to the possible outflow direction.

Three targets (including IRAS19114$+$0002) possess OH spectra which are rich in linearly 
polarized features and do not exhibit the depolarization effect (shown by small symbols in Fig. 5). 
All these objects have highly aligned fields, as judged from the polarization position angle 
profiles. In contrast, the three sources with prominent depolarization (open symbols in Fig. 5) 
exhibit greater deviations in their linear polarization PAs across their velocity profiles. 
We suggest that they have structured magnetic fields of geometry similar to that observed 
in IRAS18276$-$1431.
We speculate that the weakly polarized emission from these bright sources is intrinsically 
polarized but has been depolarized in regions of complex magnetic field. Alternatively,
intrinsically weakly or unpolarized emission due to the relationship between our viewing angle
and the source magnetic field configuration and maser beaming effects (Gray \& Field\, \cite{gray95};
Elitzur\, \cite{elitzur96,elitzur98}) becomes polarized with the PA varying steeply as 
a function of angular position and apparent velocity. As pointed out in section 5.1 the Faraday 
rotation in these objects is probably too small to affect the detectability of linear polarization.

\subsection{Prevalence of orthogonal fields}
These single dish measurements of the polarization position angle of OH masers provide 
useful information on the orientation of the magnetic field in the maser region.  
It has been commonly argued that features with linear polarization are $\sigma$ components 
of the Zeeman pattern (Garcia-Barreto et al.\,\cite{garciabarreto88}; 
Gray \& Field\,\cite{gray95}; Szymczak et al.\,\cite{szymczak98}), implying that 
the projected angle of the magnetic field on the plane of the sky is perpendicular 
to the polarization position angle. Several objects in the sample have been imaged in 
the optical and/or infrared showing axisymmetric morphology. 13 of them exhibit linearly
polarized OH maser emission, so that one can compare the position angle of the magnetic
field with that of the outflow axis. The $p$-flux-weighted average polarization angles 
for these sources are given in Table 3 for one or two OH lines together with the position 
angle of the long axis of the nebula taken from the literature. A plot of the PA of the 
mean magnetic field versus the PA of the long axis of the nebula (Fig.7) shows a scarcity 
of objects with the magnetic field aligned with the outflow. The distribution of the absolute 
difference in PA between nebula and magnetic field axes is clearly skewed; the median value 
of this distribution is 52\degr. Thus, the magnetic field is closer to orthogonal than 
parallel direction relative to the long axis although there is a large scatter.

The dominance of magnetic field components oriented across the outflow axis appears 
to support the hypothesis of magnetic collimation of bipolar lobes or jets 
(Chevalier \& Luo\,\cite{chevalier94}; Garcia-Segura et al.\,\cite{garcia99}; 
Matt et al.\,\cite{matt00}). In this model the initially weak magnetic fields emerging 
from an AGB star are wrapped up by differential rotation into toroidal tubes which
channel the fast wind in post-AGB phase producing prolate and bipolar structures.  
Recent observations provided arguments for toroidal fields in individual PN and PPN 
(Greaves\,\cite{greaves02}; Miranda et al.\,\cite{miranda01}). 
Alternatively, prevalence of the orthogonal field can be related to the source geometry. 
It is known that OH masers in PPNe are preferentially seen along the limbs of biconical 
outflows (Zijlstra et al.\, \cite{zijlstra01}). For sources with the opening angle wide 
enough the PA of the magnetic field in the maser regions could be misaligned with the nebula
long axis.  

 \begin{table*}
\caption {The $p$-flux-weighted average polarization angle of
the linearly polarized OH maser emission at 1612\,MHz and/or 1667\,MHz
(PA$_{1612}$, PA$_{1667}$) and the position angle of nebula long axis
(PA$_{\rm axis}$) taken from optical and/or infrared or radio images.}
\begin{tabular}{l c r r r r r r c }
\hline
\hline
IRAS name & Type & N & PA$_{1612}$(\degr) & $\sigma_{\rm PA_{1612}}$(\degr) &
PA$_{1667}$(\degr) & $\sigma_{\rm PA_{1667}}$(\degr) & PA$_{\rm axis}$(\degr) & Refs \\
\hline
07331$+$0021 & d &3&$-$32.3 & 4.7   &       &      & 89(30)   & 1  \\
07399$-$1435 & c &4&        &       & 48.0  &  6.7 & 23(5)    & 1,2\\
08005$-$2356 & s &2& 22.1   & 4.4   &       &      & $-$47(5) & 3  \\
16342$-$3814 & d,d &4& $-$54.8& 8.4   &$-$51.6& 5.6  & 69(3)    & 3  \\
17150$-$3224 & s,c &5& $-$72.2&32.9   &$-$37.2&17.2  & $-$33(3) & 3  \\
17393$-$3004 & s &2& $-$23.0& 6.5   &       &      & 13(3)    & 4  \\
18266$-$1239 & c &2&  50.6  & 7.2   &       &      & 66(30)   & 5  \\
18276$-$1431 & c,c &1&  47.0  & 5.0   &  77.0 &14.8  & 20(3)    & 6  \\
18450$-$0148 & c,s &3&  82.1  &18.5   &  75.3 & 3.8  & 63(2)    & 7  \\
19114$+$0002 & c,c &2& $-$84.8& 8.4   &$-$84.0& 7.3  & 21(3)    & 3  \\ 
19219$+$0947 & c &2&  $-$0.7& 4.5   &       &      & 34(5)    & 8  \\
19255$+$2123 & s &5&  64.5  & 2.5   &       &      & 27(5)    & 9,10 \\
22036$+$5306 & c &3& $-$73.8& 5.3   &       &      & 66(2)    & 11 \\ 
\hline\end{tabular}
\vskip 0.5cm
Type of spectrum: s - single peak, d - two features, c - complex;
N: number of observations used to derive PA$_{1612}$ and PA$_{1667}$; \\
References for PA$_{\rm axis}$: (1) Meixner et al.\,\cite{meixner99}; (2) Kastner et al.\,\cite{kastner98}; 
(3) Ueta et al.\,\cite{ueta00}; (4) Philipp et al.\,\cite{philipp99}; (5) Chapman\,\cite{chapman88};
(6) Bains et al.\,\cite{bains03a}; (7) Imai et al.\,\cite{imai02}; (8) Seaquist \& Davis\,\cite{seaquist83};
(9) Miranda et al.\,\cite{miranda00}; (10) Miranda et al.\,\cite{miranda01}; (11) Sahai et al.\,\cite{sahai03}  
\end{table*}

\begin{figure}
   \resizebox{\hsize}{!}{\includegraphics{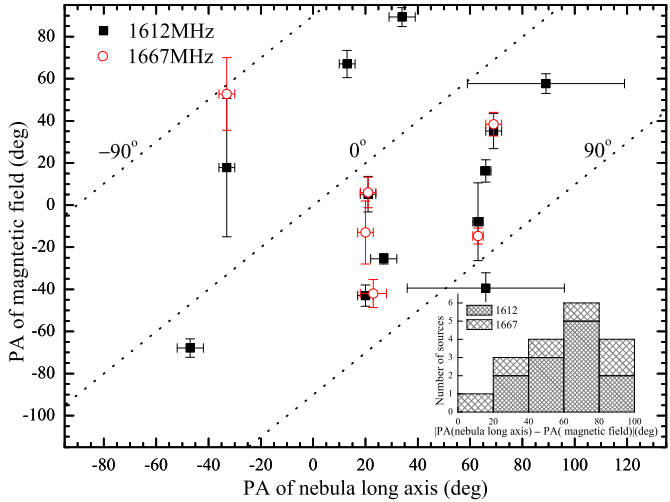}}
   \caption{Position angle of the magnetic field vector versus
position angle of the long axis of the nebula. Squares and circles
mark the 1612 and 1667\,MHz masers respectively. The diagonal 
dotted lines correspond to constant differences of $-$90, 0 
and 90\degr between the two position angles. The inset shows 
the distribution of the absolute difference in PA.}
   \label{fig7}
\end{figure}

\subsection{Disappearance of OH masers}
No OH emission was detected in 5 targets (Table 1). In IRAS06530$-$0213 weak 
1667\,MHz emission (0.22 and 0.16\,Jy) at 27.8 and 29.8\,km\,s$^{-1}$ was observed 
in 1986 and 1987 (Likkel\,\cite{likkel89}). No 1612\,MHz emission was detected 
(Likkel \cite{likkel89}; Hu et al.\,\cite{hu94}). The source was classified
recently as a carbon-rich PPN (Hrivnak \& Bacham\,\cite{hrivnak03}).
IRAS16559$-$2957 was detected at 1612\,MHz in 1986 with a 0.58\,Jy peak near 
56.2\,km\,s$^{-1}$ (Likkel\,\cite{likkel89}). At a similar epoch te Lintel 
Hekkert et al.\,(\cite{hekkert91}) observed the blue-shifted emission at 
57.0\,km\,s$^{-1}$ (0.43\,Jy) and red-shifted emission at 86.5\,km\,s$^{-1}$ 
(0.15\,Jy), while in 1991 Hu et al. (\cite{hu94}) found a 0.30\,Jy peak near 
about 70\,km\,s$^{-1}$ at 1612 and 1665\,MHz. In IRAS17079$-$3844 a 1612\,MHz 
multi-peak profile was seen in 1987 (te Lintel Hekkert et al.\,\cite{hekkertetal91}).
The 1667\,MHz maser from IRAS17436+5003 was found by Likkel (\cite{likkel89}) 
in 1986 and 1987 near a velocity of $-$26\,km\,s$^{-1}$. Recent observations 
by Bains et al. (\cite{bains03b}) suggest that the disappearance of OH maser in
this source can be permanent. In IRAS19127+1717, 1667\,MHz emission was detected 
at 15.6\,km\,s$^{-1}$ (0.24\,Jy) in 1987 (Likkel\,\cite{likkel89}). She suggested 
that the emission is probably not related to the nebula.

We point out that all 5 targets not detected in the survey were known as weak 
(0.2$-$0.4\,Jy) OH emitters about 15$-$17 years ago. Their emission has dropped 
below $\sim$0.1\,Jy. These objects need further observations to decide whether 
the emission disappeared temporarily or permanently.

 One of the most interesting objects in the sample IRAS17393$-$2727 (OH0.9+1.3)
continues a growth of the blue-shifted emission at a rate of about 1\,Jy\,yr$^{-1}$
(Shepherd et al.\,\cite{shepherd90}), whereas its red-shifted emission near 
$-$93.5\,km\,s$^{-1}$ appears to be constant, within a 0.2\,Jy accuracy, as compared
to Shepherd's data taken about 14\,yrs ago.

\section{Conclusions}
High sensitivity observations of the 1612 and 1667\,MHz OH masers from a sample 
of 47 proto-planetary nebula candidates have provided new  information on 
the polarization pattern. 

A large fraction (55\%) of the sources exhibit linearly polarized features in one 
or both OH lines. Circularly polarized features are present in 76\% of the sources.  
Generally the maser features are elliptically polarized; the degrees of linear and 
circular polarization are usually low ($<$15\%), but in some features this is as 
high as 50$-$80\%. The linearly polarized features are usually narrow 
($<$0.5\,km\,s$^{-1}$) and weak ($<$4\,Jy).

There are several sources with a large velocity extent ($>$4\,km\,s$^{-1}$) of 
linearly polarized flux for which a diversity of variations of polarization position 
angle $\chi$ across the profile is observed. The sources with large changes in 
$\chi$ are very likely to have structured magnetic fields, as recently revealed for
OH17.7$-$2.0. They also exhibit depolarization of linearly polarized emission.  
The sources with nearly constant values of $\chi$ possibly have magnetic fields 
highly aligned with specific directions and do not show depolarization effects. 
Both types of objects appear to be good candidates for future mapping of magnetic 
fields with high angular and spectral resolutions.

For the subset of PPNe with optical images showing axisymmetric morphology we found 
a dominance of magnetic field components misaligned by $>60$\degr with the long
axis of the nebulae. This finding lends support to the theoretical model of magnetic 
collimation of bipolar lobes in planetary nebulae.

The magnetic field strength inferred from likely Zeeman pairs detected in two targets 
is of the order 1$-$2\,mG; a value which is consistent with previous estimates for PPN 
and AGB stars. The upper limit of the electron density of 1\,cm$^{-1}$ in OH17.7$-$2.0 
derived from a comparison of $\chi$ angles at 1612 and 1667\,MHz lines implies a largely 
neutral envelope.

The overshoot of the 1667\,MHz velocity extent relative to the 1612\,MHz velocity extent 
is common (81\%) in the studied sources. This effect, rarely seen in AGB stars, can be 
related to changes in the source geometry and excitation conditions in the envelopes 
during the transition from AGB to PN phase.

\begin{acknowledgements}
We thank A. Richards for useful comments on an earlier version of this paper,
P. Colom for the communication of his results of polarization measurements of the NRT 
prior to publication.
The Nancay Radio Observatory is the Unit\'e Scientifique de Nancay of 
the Observatoire de Paris, associated with the CNRS. The Nancay Observatory
gratefully acknowledges the financial support of the Region Centre in France.
\end{acknowledgements}

\end{document}